\documentclass[a4paper,11pt]{article}
\usepackage{epsfig}
\usepackage{amsmath}
\usepackage{jheppub} 
\def\be{\begin{equation}}
\def\ee{\end{equation}}
\def\bea{\begin{array}}
\def\beqa{\begin{eqnarray}}
\def\eeqa{\end{eqnarray}}
\def\beqas{\begin{eqnarray*}}
\def\eeqas{\end{eqnarray*}}

\def\bp{\begin{picture}}
\def\ep{\end{picture}}
\def\bc{\begin{center}}
\def\ec{\end{center}}
\def\bfig{\begin{figure}}
\def\efig{\end{figure}}

\def\bit{\begin{itemize}}
\def\eit{\end{itemize}}

\def\[{\left[}
\def\]{\right]}
\def\({\left(}
\def\){\right)}

\def\..{\left.}
\def\.{\right.}

\def\ep{\epsilon}

\title{\boldmath A pedagogical review on muon $g-2$}
\author[a,b]{Song Li,}
\author[a,b]{Yang Xiao,}
\author[a,b]{Jin Min Yang}

\affiliation[a]{CAS Key Laboratory of Theoretical Physics, Institute of Theoretical Physics, Chinese Academy of Sciences, Beijing 100190, China}
\affiliation[b]{School of Physics Sciences, University of Chinese Academy of Sciences,  Beijing 100049, China}

\emailAdd{lisong@itp.ac.cn}
\emailAdd{xiaoyang@itp.ac.cn}
\emailAdd{jmyang@itp.ac.cn}

\abstract{This note is a pedagogical mini review on the muon anomalous magnetic moment 
($g-2$), translated and adapted from our article published in Modern Physics 4 (2021) 40-47. 
The contents include: 
(i) The magnetic moment of an electric-current coil;
(ii) The  magnetic moment of a charged lepton estimated as a classical charged ball with spin;          
(iii) The  magnetic moment of a charged lepton from Dirac equation with electromagnetic interaction; 
(iv) The $g-2$ of a charged lepton from QED beyond tree level with effective couplings;
(v) The measurement of muon $g-2$;
(vi) The muon $g-2$ in low energy supersymmetric models.
Finally, we give an outlook.}    

\begin{document}
\maketitle \indent
\newpage
\section{Introduction}

Although the phenomenological success of the Standard Model (SM) is tremendous, especially 
the discovery of the Higgs boson completed the list of all particles predicted by the SM,  
there are still some unanswered questions, such as the asymmetry between matter and 
anti-matter, the hierarchy problem, and the dark matter mystery. Therefore, searching for 
new physics beyond the SM is the main theme in today's particle physics. 

Since particle physics is a discipline relying on experiments, the experimental crises 
or deviations from the current paradigm theory play a crucial role in the pursue of new 
physics. 
This April the Fermilab announced its first measurement result \cite{FNAL:gmuon} 
on the muon anomalous magnetic dipole 
moment ($g-2$), which, combined with the BNL result \cite{BNL:gmuon}, shows a $4.2\sigma$ deviation from the 
SM prediction \cite{Aoyama:2020ynm}. 
This greatly enhanced 
the confidence of particle physicists in probing new physics beyond the SM.
Utilizing this result, we may know what new theories are favored or excluded. 
So such a muon $g-2$ anomaly is extremely important in particle physics.  

Given the important role played by the muon $g-2$, we in this note give a pedagogical mini 
review to help those beginners to grasp those knowledge quickly.
Starting from the magnetic moment of an electric-current coil and the estimation 
of the  magnetic moment of a lepton as a classical charged ball with spin,          
we derive the  magnetic moment of a lepton from Dirac equation with electromagnetic 
interaction.     
Then we dicusss the anomalous magnetic moment ($g-2$) of a lepton from QED beyond tree level 
with effective couplings. 
After describing the measurement of muon $g-2$, we discuss the explanation of the muon $g-2$ 
anomaly in low energy supersymmetric models.
Finally, we give an outlook.   

\section{The \texorpdfstring{$g-2$}{g-2} of a charged lepton}
In this section we will start from the original definition of magnetic moment of an electric-current coil 
in electromagnetism. 
For the magnetic moment (non-anomalous) of a charged lepton, we will first delineate it assuming the lepton 
as a classical charged rigid-body and then derive it from both the Dirac equation and the tree-level QED. 
Finally we derive the $g-2$ of a charged lepton from the QED beyond tree level. 
\subsection{The magnetic moment of an electric-current coil}
\begin{figure}[htb]
\begin{center}
\includegraphics[width=6cm]{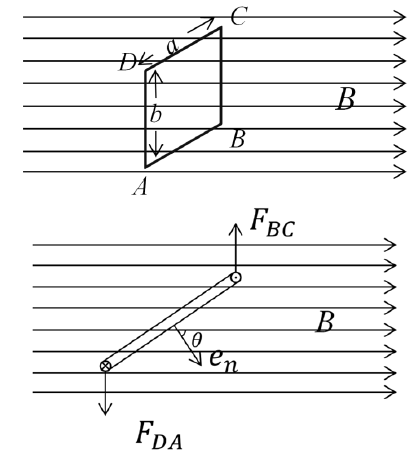}
\end{center}
\vspace{-.5cm}
\caption{A rectangular electric-current coil in an uniform magnetic field.}
\label{fig1}
\end{figure}

Consider a rectangular electric-current coil in an uniform magnetic field, 
as shown in Fig.\ref{fig1}.
From Ampere's law we know the net force on the coil is zero. But the moment of force on the 
coil is not zero, whose magnitude is given by 
\begin{equation}
L=F_{BC} \frac{a}{2}\sin\theta+F_{DA} \frac{a}{2}\sin\theta=IabB\sin\theta . 
\end{equation} 
Considering the direction, the moment of force takes the form 
\begin{equation}
\vec{L}=IA\vec{e}_n \times \vec{B}=\vec{M}\times \vec{B}, 
\end{equation}
where $A$ is the coil area, $\vec{e}_n$ is the unit vector of the normal direction of the coil, 
and the magnetic moment  $\vec{M}$ is defined as 
 \begin{equation}
 \vec{M} =IA\vec{e}_n .
\end{equation}
It is proved in electromagnetism \cite{zhao2018ele} that the above two formulas apply to  
planar coils of arbitrary shape.
    
\subsection{The magnetic moment of a charged lepton as a classical rigid-body}          
If a charged lepton like a muon is regarded as a charged rigid body, and assuming that its charge density $\rho_e$ is proportional to the mass density $\rho_m$, that is $\rho_e=\alpha\rho_m$, and thus $e =\alpha m$, where $e$ and $m$ are the charge and mass of the lepton, respectively. The spin that the lepton carries corresponds to a rotation in classical mechanics. If this charged rigid body rotates around the $z$-axis with an angular velocity $\omega$, then its magnetic moment can be obtained by an integration:
\begin{equation}
    \begin{aligned}
        M &= \int A\,{\rm d}I\\
        &= \int\pi\left(x^2+y^2\right)\frac{\rho_e(\vec{r}~){\rm d}V }{2\pi/\omega} \\
        &= \frac{\alpha}{2}\left(\omega\int\left(x^2+y^2\right)\rho_m(\vec{r}~){\rm d}V \right)\\
        &= \frac{\alpha}{2}L = \frac{e}{2m}L,
    \end{aligned}
\end{equation}
 where $L$ is the rotational angular momentum of this rigid body. Writing the above result in vector form, we obtain  
\begin{equation}\label{eq:classcalMDM}
    \vec{M}= \frac{e}{2m} \vec{L}.  
\end{equation}

If this magnetic moment is intrinsic, then its magnitude will not change, and its potential energy in the magnetic field is 
\begin{equation}\label{eq:MDMenergy}
    E=-\vec{M}\cdot\vec{B},
\end{equation}
where $\vec{B}$ is the magnetic induction intensity. 


Of course, such a classical way cannot correctly describe the property of a charged lepton
and Eq.~(\ref{eq:classcalMDM}) needs to be re-derived from Dirac equation with electromagnetic 
interaction or from QED.   

\subsection{The magnetic moment of a charged lepton from Dirac equation}
In the following we derive such a potential in Eq.(\ref{eq:MDMenergy}) from  Dirac equation with electromagnetic 
interaction~\cite{zeng2007quantum}. A charged lepton is described by a Dirac spinor wave function $\psi(x)$ 
which satisfies Dirac equation. With electromagnetic interaction, the Dirac equation
takes the form  
\begin{equation}\label{eq:diraceq}
    i\frac{\partial}{\partial t}\psi=\left[\vec{\alpha}\cdot\left(\vec{P}-\frac{e}{c}\vec{A}\right)+e\phi+mc^2\beta\right]\psi,
\end{equation}
where $\phi$ is the electric potential and $c$ is the speed of light. We take 
$\hbar=1$ but keep $c$ for the convenience to make a non-relativistic approximation in the following derivation. For $c=1$ the above equation can reduce to the form 
$\left[ i\gamma^\mu(\partial_\mu+ieA_\mu) -m\right]\psi=0$ in quantum field theory.
The relationships between the matrices $\vec{\alpha}$, $\beta$ and the Dirac $\gamma$ are $\beta=(\gamma^0)^{-1}$ and $\alpha^i=\beta \gamma^i$. In Pauli-Dirac representation, we have
\begin{align}
    \beta = \begin{pmatrix}
    I & 0\\
    0 & -I
    \end{pmatrix}, ~~~~
    \alpha^i = \begin{pmatrix}
    0 & \sigma^i\\
    \sigma^i & 0
    \end{pmatrix}.
\end{align}
We express the four-component wave function as two two-component wave functions:
\begin{equation}\label{eq:two-spinor}
    \psi=\begin{pmatrix}
    \varphi\\
    \chi
    \end{pmatrix}{\rm e}^{-imc^2t},
\end{equation}
where the static energy of the lepton has been separated. Substitute Eq.~(\ref{eq:two-spinor}) into Eq.~(\ref{eq:diraceq}), we obtain
\begin{align}
    i\frac{\partial}{\partial t}\varphi &= c\vec{\sigma}\cdot\left(\vec{P}-\frac{e}{c}\vec{A}\right)\chi+e\phi\varphi, \label{eq:two-spin-eq01}\\
    i\frac{\partial}{\partial t}\chi &= c\vec{\sigma}\cdot\left(\vec{P}-\frac{e}{c}\vec{A}\right)\varphi+e\phi\chi-2mc^2\chi. \label{eq:two-spin-eq02}
\end{align}
In the non-relativistic limit, the terms that do not contain $c$ in Eq.~(\ref{eq:two-spin-eq02}) can be ignored, and thus we have 
\begin{equation}
    \chi\approx\frac{1}{2mc}\vec{\sigma}\cdot\left(\vec{P}-\frac{e}{c}\vec{A}\right)\varphi.
\end{equation}
Substituting this result into Eq.~(\ref{eq:two-spin-eq01}) and taking $c=1$, we obtain
\begin{equation}\label{eq:sub-pauli-eq}
    i\frac{\partial}{\partial t}\varphi = \frac{1}{2m}\left[\vec{\sigma}\cdot\left(\vec{P}-e\vec{A}\right)\right]^2\varphi+e\phi\varphi.
\end{equation}
Using the relation
\begin{equation}
    (\vec{\sigma}\cdot\vec{a})(\vec{\sigma}\cdot\vec{b})=\vec{a}\cdot\vec{b}+i\vec{\sigma}\cdot\left(\vec{a}\times\vec{b}\right),
\end{equation}
we can get
\begin{equation}\label{eq:simplify-sigma}
    \begin{aligned}
        \left[\vec{\sigma}\cdot\left(\vec{P}-e\vec{A}\right)\right]^2 &= \left(\vec{P}-e\vec{A}\right)^2 + i\vec{\sigma}\cdot\left[\left(\vec{P}-e\vec{A}\right)\times\left(\vec{P}-e\vec{A}\right)\right]\\
        &= \left(\vec{P}-e\vec{A}\right)^2 - ie\vec{\sigma}\cdot\left[\vec{P}\times\vec{A}+\vec{A}\times\vec{P}\right]\\
        &= \left(\vec{P}-e\vec{A}\right)^2 - e\vec{\sigma}\cdot(\nabla\times\vec{A})\\
        &= \left(\vec{P}-e\vec{A}\right)^2 - e\vec{\sigma}\cdot\vec{B}.
    \end{aligned}
\end{equation}
Substituting Eq.~(\ref{eq:simplify-sigma}) into Eq.~(\ref{eq:sub-pauli-eq}), we have
\begin{equation}
    i\frac{\partial}{\partial t}\varphi = \left[\frac{1}{2m}\left(\vec{P}-e\vec{A}\right)^2-\frac{e}{m}\frac{\vec{\sigma}}{2}\cdot\vec{B}+e\phi\right]\varphi,
\end{equation}
where $\vec{\sigma}/2$ is the spin $\vec{S}$ of the lepton. Therefore, in non-relativistic approximation,
\begin{equation}
    H = \frac{1}{2m}\left(\vec{P}-e\vec{A}\right)^2-\frac{e}{m}\vec{S}\cdot\vec{B}+e\phi.
\end{equation}
Now comparing this result with Eq.~(\ref{eq:MDMenergy}), we obtain
\begin{equation}
    \vec{M}=\frac{e}{m} \vec{S}=g\frac{e}{2m} \vec{S}  ~~~~~~(g=2),
\end{equation}
where the factor $g$ is called the Landé $g$ factor.

\subsection{The magnetic moment of a charged lepton from QED at tree level}  
The potential of a charged lepton with a  magnetic moment $\vec{M}$ 
in magnetic field $\vec{B}$ takes a form in Eq.(\ref{eq:MDMenergy}).
So the  magnetic moment of  a charged lepton can be found out from such a potential.  

\begin{figure}[htb]
\begin{center}
\includegraphics[width=6cm]{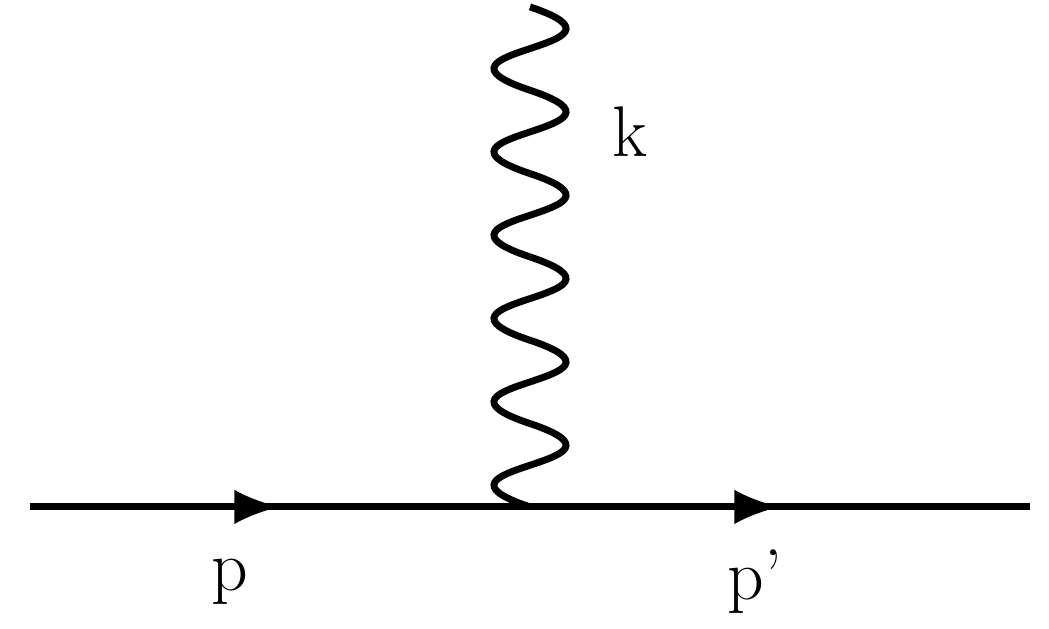}
\end{center}
\vspace{-.5cm}
\caption{The tree level Feynman diagram showing the interaction of a charged lepton 
with the electromagnetic field.  The interaction vertex is $e\gamma^\mu$. }
\label{fig3}
\end{figure}

In QED, as depicted in Fig.\ref{fig3}, the electromagnetic interaction of a charged lepton 
takes a form at tree level 
\begin{equation}
{\cal H_I} = - e A_\mu(x) \bar\psi(x) \gamma^\mu \psi(x),
\end{equation}
where $ A_\mu(x) $ is the four-vector potential of the electromagnetic field,
$e$ is the magnitude of the electric charge of the charged lepton and $\gamma^\mu$ is the  
Dirac algebra representation matrix given by 
\begin{equation}
\gamma^\mu = \left( \begin{array}{cc} 0 & \sigma^\mu \\ \bar{\sigma}^\mu & 0 \end{array} \right) ,
\end{equation}
with  $\sigma^\mu$ being the Pauli matrices. 

Since we only consider the influence of the applied magnetic field, we have 
$ A_\mu(x) =\left( 0, -\vec{A}(\vec{x})\right)$ and 
\begin{equation}
{\cal H_I} = e A^i(x) \bar\psi(x) \gamma^i \psi(x).
\end{equation}  
In Weyl representation we obtain the free plane-wave solution of the Dirac equation
of a charged lepton 
(we omit the plane-wave factor $e^{-ipx}$ for the time being) \cite{peskin2018introduction}
\begin{equation}
u(p)= \frac{1}{\sqrt{2E_p}}  \left( \begin{array}{c}  \sqrt{p\cdot \sigma} \xi \\
                             \sqrt{p\cdot \bar\sigma} \xi  \end{array} \right)
\simeq \frac{1}{\sqrt{2}}  \left(  \begin{array}{c} (1-\vec{p}\cdot \vec{\sigma}/2m)\xi \\
                             (1+\vec{p}\cdot \vec{\sigma}/2m)\xi  \end{array} \right)
        +{\cal O}(|\vec{p}|^2),  
\end{equation}  
 where $m$ is the mass of the charged lepton, 
and the normalization constant $1/\sqrt{2E_p}$ is to make $u^\dagger u=1$. 
So we have 
\begin{equation}
\bar{u}(p') \gamma^i u(p) \simeq \xi'^\dagger 
\left( \frac{\vec{p'}\cdot \vec{\sigma}}{2m}\sigma^i
+\sigma^i  \frac{\vec{p}\cdot \vec{\sigma}}{2m} \right)\xi .
\end{equation}
From the property of the Pauli matrices 
\begin{eqnarray}
&& \left[ \sigma^i, \sigma^j\right]= 2i\epsilon^{ijk} \sigma_k, \\
&& \left\{ \sigma^i, \sigma^j\right\}= 2\delta^{ij} I, \\
&& 2 \sigma^i \sigma^j=  \left[ \sigma^i, \sigma^j\right]+ \left\{ \sigma^i, \sigma^j\right\},
\end{eqnarray}
we obtain
\begin{equation}
\bar{u}(p') \gamma^i u(p) \simeq 
\frac{(p+p')^i}{2m}  \xi'^\dagger  \xi 
+ \left(\frac{-i}{m}\epsilon^{ijk}k^j\right)  \xi'^\dagger \frac{\sigma^k}{2} \xi,
\end{equation}
where $k^i=p'^i-p^i$. 
Herefater we will not consider the first term which does not contain $\sigma$ matrices 
and thus is irrelevant to spin and magnetic moment.
In the second term, $ip'^i$ and  $-ip^i$ come from $\partial^i\bar\psi$ and $\partial^i\psi$,
respectively. 
Retaining the plane-wave factor $e^{-ipx}$, we have 
\begin{eqnarray}
&& \left(\frac{-i}{m}\epsilon^{ijk}k^j\right)  \xi'^\dagger \frac{\sigma^k}{2} \xi  \nonumber\\ 
&&\Rightarrow 
eA^i(x) \frac{-\epsilon^{ijk}}{m} 
\left[  \partial^j(\xi'e^{-ip'x})^\dagger   \frac{\sigma^k}{2} (\xi e^{-ipx})
+(\xi'e^{-ip'x})^\dagger \frac{\sigma^k}{2}  \partial^j(\xi e^{-ipx}) \right] \nonumber\\
&& = -\frac{e}{m} A^i(x) \epsilon^{ijk}  \partial^j 
\left[ (\xi'e^{-ip'x})^\dagger  \frac{\sigma^k}{2} (\xi e^{-ipx}) \right]  \nonumber\\
 && = -\frac{e}{m} \epsilon^{kji} \partial^j\left(A^i(x)\right) 
\left[ (\xi'e^{-ip'x})^\dagger  \frac{\sigma^k}{2} (\xi e^{-ipx}) \right]+ {\rm total ~derivative ~term},   
\label{5-11}  
\end{eqnarray}
where the total derivative term can be neglected. 
Since the magnectic potential $A^i(x)$ and the magnetic field $B^i(x)$ are related as 
 \begin{equation} \label{5-12} 
B^k = \epsilon^{kji} \partial^j A^i,
\end{equation}
then we have the spin-dependent (SD) interaction 
 \begin{equation} \label{5-13} 
{\cal H_I}(SD)= - \vec{B}\cdot \left(\frac{e}{m} \vec{S}\right) 
              = - \vec{B}\cdot \vec{M}, 
\end{equation}
with $\vec{M}$ being the magnetic moment of a charged lepton 
 \begin{equation} \label{5-14} 
\vec{M}=  \frac{e}{m} \vec{S}=g\frac{e}{2m} \vec{S}  ~~~~~~(g=2) ,
\end{equation}
and $\vec{S}$ being the spin of a charged lepton (see eq.(3.111) in \cite{peskin2018introduction})  
 \begin{equation}
\vec{S}  =  (\xi'e^{-ip'x})^\dagger  \frac{\vec\sigma}{2} (\xi e^{-ipx}) .
\end{equation}

\subsection{The \texorpdfstring{$g-2$}{g-2} of a charged lepton from QED beyond tree level}
\begin{figure}[htb]
\begin{center}
\includegraphics[width=6cm]{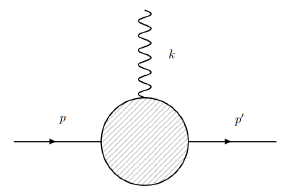}
\end{center}
\vspace{-.5cm}
\caption{The Feynman diagram showing the interaction of a charged lepton 
with the electromagnetic field beyond tree level with an effective vertex. }
\label{fig4}
\end{figure}

In QED beyond tree level, as shown in Fig.\ref{fig4},  
the interaction vertex of a charged lepton with 
electromagnetic field takes the effective form   
 \begin{equation}
-ie \epsilon_\mu(k) \bar{u}(p') \Gamma^\mu u(p), 
\end{equation}
where $\epsilon_\mu(k)$ is the polarization vector of the electromagnetic field and 
$  \Gamma^\mu$ is an effective form invloving all Lorentz vectors in Fig.\ref{fig4}.  

The general form of $\Gamma^\mu$ includes  $\gamma^\mu$, $k^\mu=p'^\mu-p^\mu$,  $P^\mu=p'^\mu+p^\mu$,
and some contractions of the anti-symmetric tensor $\epsilon^{\mu\nu\alpha\beta}$ with the momentums.
Utilizing the relations 
\begin{eqnarray}
&& \epsilon^{\mu\nu\alpha\beta} =i\gamma^{[\mu} \gamma^\nu  \gamma^\alpha \gamma^{\beta]} \gamma^5,\\ 
&& \gamma^\mu p_\mu u(p)=m u(p), \\
&& \bar{u}(p) \gamma^\mu p_\mu = \bar{u}(p) m,  \\
&& \gamma^\mu \gamma^\nu +\gamma^\nu   \gamma^\mu =2 g^{\mu\nu},
\end{eqnarray}
we can reform  $\Gamma^\mu$ not to contain the contrations of $\gamma$ matrics or 
$\epsilon^{\mu\nu\alpha\beta}$ with the momentums.
Finally,  $\Gamma^\mu$  takes a form 
  \begin{equation}
 \Gamma^\mu =\gamma^\mu A_1 +\frac{P^\mu}{2m}A_2 +i\frac{k^\mu}{2m}A_3 + \gamma^\mu \gamma_5 A_4
            +\frac{k^\mu}{2m} \gamma_5 A_5 + i\frac{P^\mu}{2m}\gamma_5  A_6,  
\end{equation}
where the factor $1/2m$ ensures the corresponding $A's$ to be dimensionless, and the factor $i$ 
ensures the corresponding $A's$ to be real so that $\epsilon_\mu(k) \bar{u}(p') \Gamma^\mu u(p)$ 
is hermitian.   
As the functions of $k^2$, the coefficients $A_i$ are Lorentz scalars.

Since $\bar{u}(p') \Gamma^\mu u(p)$ is a conserved current in QED, we have 
 \begin{equation}
k_\mu \bar{u}(p') \Gamma^\mu u(p)=0 ,
\end{equation}
which will further restrain the form of $\Gamma^\mu$. 
From the relations 
\begin{eqnarray}
&& k_\mu P^\mu =0, \\
&& k_\mu \bar{u}(p') \gamma^\mu u(p)=0, \\
&& k_\mu \bar{u}(p') \gamma^\mu \gamma_5 u(p)=2m \bar{u}(p') \gamma_5 u(p),
\end{eqnarray}
we have 
 \begin{equation}
k_\mu \bar{u}(p') \Gamma^\mu u(p)= \bar{u}(p') 
\left( i\frac{k^\mu k_\mu}{2m}A_3 + 2m \gamma_5 A_4
            +\frac{k^\mu k_\mu}{2m} \gamma_5 A_5 \right) u(p)=0,  
\end{equation}
which leads to $A_3=0$ and $A_5=-A_4 4m^2/k^2$. Then $\Gamma^\mu $ takes a form 
 \begin{equation}
 \Gamma^\mu =\gamma^\mu A_1 +\frac{P^\mu}{2m}A_2 
+\left(\gamma^\mu-\frac{2mk^\mu}{k^2}\right) \gamma_5 A_4 
+ i\frac{P^\mu}{2m}\gamma_5  A_6.  
\end{equation}
Further, utilizing the Gordon identities 
\begin{eqnarray}
&& \bar{u}(p') \frac{P^\mu}{2m} u(p)=
 \bar{u}(p')\left(\gamma^\mu-i\sigma^{\mu\nu} \frac{k_\nu}{2m}\right)  u(p), \\
&& \bar{u}(p') \frac{P^\mu}{2m}\gamma_5 u(p)=
 \bar{u}(p')\left(-i\sigma^{\mu\nu} \frac{k_\nu}{2m}\gamma_5 \right)  u(p), 
\end{eqnarray}
we have 
 \begin{equation} \label{6-15}
 \Gamma^\mu =\gamma^\mu F_E(k^2) 
+\left(\gamma^\mu-\frac{2mk^\mu}{k^2}\right) \gamma_5 F_A(k^2)  
+ i\sigma^{\mu\nu} \frac{k_\nu}{2m} F_M(k^2)
+ \sigma^{\mu\nu} \frac{k_\nu}{2m}\gamma_5 F_D(k^2), 
\end{equation}
where the renamed $F$ coefficients are called form factors. 

Since the parameter $e$ in the Hamiltonian density will be corrected by 
the $F$  form factors, it is no longer the measured electric charge which 
needs to be redefined. For this end, we take $A_\mu=(\phi(x),\vec{0})$
and then only need to consider $\Gamma^0$. Taking the zero-momentum limit 
and utilizing 
  \begin{equation} 
  u(p)\simeq \frac{1}{\sqrt{2}} \left( \begin{array}{c} \xi \\ \xi \end{array} \right) 
 +{\cal O}(\vec{p}),
  \end{equation}  
and 
 \begin{equation} 
\gamma^0 \gamma^0 = \left( \begin{array}{cc} ~~1~~ & ~~0~~  \\ 0 & 1 \end{array} \right),
~~~~~~~\gamma^0 \gamma^0 \gamma_5 = 
       \left( \begin{array}{cc} ~~-1~~ & ~~0~~  \\ 0 & 1 \end{array}  \right),
  \end{equation} 
we obtain 
 \begin{equation}
\bar{u}(p') \Gamma^0 u(p) \simeq F_E(0) \xi^\dagger \xi + {\cal O}(\vec{p},\vec{p'}) , 
 \end{equation}
which means we have a term in the Hamiltonian density:
 \begin{equation}
e F_E(0) \phi(x) \left[ (\xi' e^{-ip'x})^\dagger (\xi e^{-ipx}) \right].
 \end{equation}
This term is just the electric potential and thus $e F_E(0)$ is the physical charge.
So we need to redefine the electric charge to be $e$, which means  $F_E(0)=1$. This 
is called the renormalization condition of the electric charge.
Therefore, the first term of $\Gamma^\mu$ is $\gamma^\mu+ {\cal O}(k^2)$. 

Since we are concerned primarily with the magnetic moment, we focus on the $F_M$ term.
For simplicity we set $A_\mu=\left( 0,\vec{A}(\vec{x})\right)$.
In the low energy limit $k_\nu=(0,-\vec{k})$, for $\sigma^{\mu\nu}$ only the spatial 
components  $\sigma^{ij}$ need to be considered. 
On the other hand, since  the $F_M$ term already contains the momentums to the first power,
we can use the zero-momentum limit of $u(p)$.
Also noticing the relation
 \begin{equation}
\gamma^0 \sigma^{ij} = \epsilon^{ijk} \left( \begin{array}{cc} ~~0~~&~~\sigma^k~~ \\ 
 \sigma^k & 0 \end{array} \right), 
 \end{equation}
we have 
 \begin{equation}   
\bar{u}(p') i\sigma^{i\nu} \frac{k_\nu}{2m} F_M(k^2) u(p) \simeq 
\left( \frac{-iF_M(0)}{m}  \epsilon^{ijk}k^j \right) \xi'^\dagger \frac{\sigma^k}{2}\xi .
\end{equation}
Compared with eqs.(\ref{5-11}-\ref{5-13}), this leads to an additional term in the Hamiltonian
 \begin{equation}  
 -\frac{eF_M(0)}{m} \vec{B}\cdot  
\left[ (\xi'e^{-ip'x})^\dagger  \frac{\vec{\sigma}}{2} (\xi e^{-ipx}) \right].
\end{equation} 
Compared with eq.(\ref{5-14}), we have
 \begin{equation} 
\vec{M}=g\frac{e}{2m} \vec{S} ,
\end{equation}
where 
\begin{eqnarray}
&& g=2+2F_M(0), \\
&& a\equiv \frac{g-2}{2}=F_M(0). 
\end{eqnarray}
In the calculation of $F_M(0)$, we usually need not to derive the whole $\Gamma^\mu$.
We can use the $\gamma$ algebra to construct a model-independent projection matrix
$P_M$ and then calculate $tr( \Gamma^\mu P_M)$.

Before ending this section, we digress a little to discuss the form of electric dipole moment
of a charged lepton. 
For this end, consider the $F_D(k^2)$ term in eq.(\ref{6-15}) in the low energy limit and 
set $A_\mu =\left( \phi(x), \vec{0} \right)$. Then for  $\sigma^{\mu\nu}$  
we only need to cosider  $\sigma^{0i}$.
Noticing the relation   
 \begin{equation} 
\gamma^0 \sigma^{0i} \gamma_5 = 
       i \left( \begin{array}{cc} ~~0~~ & ~~\sigma^i~~  \\ \sigma^i  & 0 \end{array}  \right),
  \end{equation} 
we have 
\begin{eqnarray}   
\bar{u}(p') \sigma^{\mu\nu} \frac{k_\nu}{2m} F_D(k^2) u(p) 
  &\simeq &
  \left( \frac{-iF_D(0)}{m} k^i \right) \xi'^\dagger \frac{\sigma^i}{2}\xi \nonumber \\
 &\Rightarrow &
  e \phi(x) \left( \frac{-F_D(0)}{m}\right)  ik^i   
\left[ \left(\xi' e^{-ip'x}\right)^\dagger \frac{\sigma^i}{2} \left(\xi e^{-ipx}\right) \right] \nonumber \\
&=&
   -\frac{e F_D(0)}{m} \phi(x) \partial^i  
\left[ \left(\xi' e^{-ip'x}\right)^\dagger \frac{\sigma^i}{2} \left(\xi e^{-ipx}\right) \right] \nonumber \\
&=&
   \frac{e F_D(0)}{m}  \left(\partial^i \phi(x)\right) 
\left[ \left(\xi' e^{-ip'x}\right)^\dagger \frac{\sigma^i}{2} \left(\xi e^{-ipx}\right) \right]  \nonumber \\
&=&
 \frac{e F_D(0)}{m} \vec{E}\cdot  \vec{S},
\end{eqnarray}
where $E^i=\partial^i \phi$.  Compared with the potential of an electric dipole momemnt in an electric
field $V(x)=- \vec{d}\cdot  \vec{E}$, we obtain the electric dipole moment of a charged lepton
 \begin{equation} 
\vec{d}= -\frac{e F_D(0)}{m} \vec{S}.
\end{equation}
 
\section{The measurement of muon \texorpdfstring{$g-2$}{g-2} at BNL and Fermilab}
So far our discussions on magnetic moment and $g-2\equiv a_\ell$ are applicable to any charged 
lepton, which in the SM can be an electron, a muon or a tau. 
If new physics at some high energy scale $\Lambda$ contributes to $a_\ell$, its contribution 
 $\delta a_\ell$ satisfies 
\begin{equation} 
\frac{\delta a_\ell}{ a_\ell} \sim \left( \frac{m_\ell}{\Lambda} \right)^2 .
\end{equation}   
This means that the $g-2$ of a heavier lepton is more sensitive to new physics. 
Although the tau lepton is heaviest, its liftime is too short and its $g-2$ is very hard to
precisely measure. The electron is stable and can be copiously produced, but its mass is too
small, about $1/200$ of the muon mass. Overall, among the leptons, the muon $g-2$ is the best
probe of new physics.   

\subsection{A description of the measuring method} 
\begin{figure}[htb]
\begin{center}
\includegraphics[width=8cm]{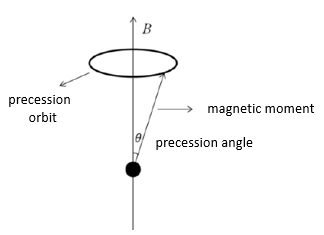}
\end{center}
\vspace{-.5cm}
\caption{The Larmor precession of a muon. }
\label{lamor}
\end{figure}

Due to its magnetic moment $\vec{M}$, an electric-current coil in a magnetic field $\vec{B}$ may have 
a non-zero moment of force  $ \vec{L}=\vec{M}\times  \vec{B}$ and hence rotate. 
For a muon, if assumed to be a uniformly charged ball, its spin may lead to 
a non-zero moment of force which make the spin rotate around the applied magnetic field:
 \begin{equation} 
\frac{d\vec{S}}{dt}=\vec{L}= \vec{M}\times  \vec{B}=g \frac{e }{2m} \vec{S}\times  \vec{B}.
\end{equation} 
 This is called the Larmor precession of a muon, as shown in Fig.\ref{lamor}. 
The angular velocity of the precession is 
 \begin{equation} 
\omega=g \frac{e }{2m} B. 
\end{equation}
So we can obtain the value of $g$ from the measurement of the muon spin-change angle in a given 
period of time. The short lifetime (2.2 $\mu s$) of a muon can be prolonged by its high velocity
from special theory of reletivity. On the other hand, technically the moving muons are easier to
control than stationary muons. So in experiments the muons are produced in some way and then 
injected into a storage ring  where the spin precessions are measured. 

Of course, in experiments the spin changes of the muons are not measured directly.
Instead, they are performed via the measurement of its decay products, i.e., the electrons.
The technical details are beyond the scope of this review. However, we should note that 
the $g-2$ measurement mentioned above is discussed assuming the muons to be stationary.                 
In real experiments, the muons are moving in a storage 
ring. So we need to perform a Lorentz transformation to obtain the formulas in the 
laboratory frame. The result in the the laboratory frame is given by \cite{book:Jegerlehner:2017}
 \begin{equation} 
\frac{d\vec{S}}{dt}= (\vec{\omega}_c+\vec{\omega}_a) \times  \vec{S},
\end{equation} 
where 
\begin{eqnarray}
&& \vec{\omega}_c=-\frac{e }{\gamma m} \left( \vec{B}+ \frac{\gamma^2 }{\gamma^2-1} 
\frac{\vec{E}\times \vec{v}}{c^2} \right),\\
&& \vec{\omega}_a=-\frac{e }{m} \left[ a\vec{B}-a \left( \frac{\gamma}{\gamma+1} \frac{\vec{v}\cdot
 \vec{B}}{c^2} \right) \vec{v}+\left(a-\frac{1}{\gamma^2-1} \right) 
\frac{\vec{E}\times \vec{v}}{c^2} \right],
\end{eqnarray}
where $v$ is the velocity of the muon and $\gamma=1/\sqrt{1-v^2/c^2}$. 
If the magnetic field is precisely perpendicular to the storage ring and the velocity
of the muon satisfies 
 \begin{equation}
a=\frac{1}{\gamma^2-1},
\end{equation} 
then $\gamma\simeq 29.3$ is called the magic $\gamma$-factor and $\vec{\omega}_a$ can be 
simplified as 
 \begin{equation}
 \vec{\omega}_a=-a \frac{e }{m} \vec{B}.
 \end{equation} 

\subsection{The results of the measurements} 
The BNL E821 result is \cite{BNL:gmuon}
 \begin{equation}
 a^{\rm BNL}_\mu= 116 592 091(63)\times 10^{-11}.
 \end{equation} 
 The Fermilab result combined with the BNL result is \cite{FNAL:gmuon}
 \begin{equation}
 a^{\rm Exp}_\mu=a^{\rm BNL+FNAL}_\mu= 116 592 061(41)\times 10^{-11}.
 \end{equation} 
The SM predition  
\begin{equation}
 a^{\rm SM}_\mu = 116 591 810(43)\times 10^{-11},
 \end{equation}
consists of the following contributions \cite{Aoyama:2012wk,Aoyama:2019ryr,Czarnecki:2002nt,Gnendiger:2013pva,Davier:2017zfy,Keshavarzi:2018mgv,Colangelo:2018mtw,Hoferichter:2019gzf,Davier:2019can,Keshavarzi:2019abf,Kurz:2014wya,Melnikov:2003xd,Masjuan:2017tvw,Colangelo:2017fiz,Hoferichter:2018kwz,Gerardin:2019vio,Bijnens:2019ghy,Colangelo:2019uex,Blum:2019ugy,Colangelo:2014qya}
\begin{eqnarray}
&& a^{\rm QED}_\mu = 116 584 718.9(1)\times 10^{-11},\\
&& a^{\rm EW}_\mu= 153.6(1)\times 10^{-11},\\
&& a^{\rm HVP, ~LO}_\mu= 6931(40)\times 10^{-11},\\
&& a^{\rm HVP, ~NLO}_\mu= -98.3(7)\times 10^{-11},\\
&& a^{\rm HVP, ~NNLO}_\mu= 12.4(1)\times 10^{-11},\\
&& a^{\rm HLBL}_\mu+a^{\rm HLBL,~NLO}_\mu= 92(18)\times 10^{-11},
\end{eqnarray}
where the main uncertainties come from the hadronic contributions in 
HVP and HLBL.
The deviation between the experiment and SM 
is   
\begin{equation}
 a^{\rm Exp}_\mu-a^{\rm SM}_\mu = 251(59) \times 10^{-11},
 \end{equation}
which is $4.2\sigma$. 
However, if the lattice simulation result of the hadronic contribution from the BMW group
 is taken,  the deviation between the experiment and SM can be reduced to  $1.5\sigma$ \cite{gm2-lattice}.

\subsection{The implication for low energy supersymmetry}
The deviation between the experiment and the SM prediction for the muon $g-2$ 
may be a harpinger of new physics beyond the SM. Among the new physics theories,
the low energy supersymmetry (SUSY) is the most popular candidate. In SUSY the smuons or 
muon sneutrino plus electroweakinos (electroweak gauginos and higgsinos) contribute to
 muon $g-2$ at one loop level, the same level as the SM. 
For a common sparticle mass $M_{SUSY}$, the SUSY contribution to  muon $g-2$ can be 
approximated as \cite{Moroi:1995yh}
\begin{equation}
 \delta a^{\rm SUSY}_\mu \sim \tan\beta \left( \frac{100 {\rm GeV}}{M_{SUSY}}\right)^2  10^{-10},
 \end{equation}   
where $\tan\beta$ is the ratio of the vacuum expectation values of the two Higgs doublets.
Clearly, to generate the required contribution to explain the muon $g-2$ deviation, a low
SUSY mass and a large $\tan\beta$ are favored.  Combined with other experimental constraints,
such as the dark matter detections and the LHC searches of sparticles, the favored parameter 
space of SUSY can be specified, which can be of some guidance for the future search of 
spartilces at the HL-LHC. 
 
\begin{figure}[htb]
\begin{center}
\includegraphics[width=11cm]{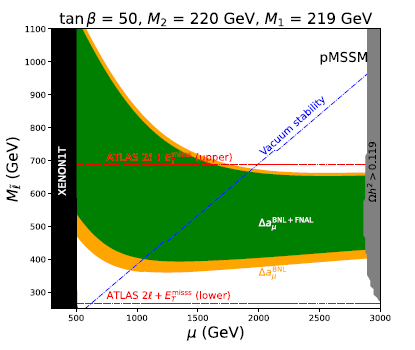}
\end{center}
\vspace{-.5cm}
\caption{ The MSSM parameter space to explain  the muon $g-2$ at $2\sigma$ 
level \cite{gm2-gutsusy-01}. The dark green region and the region 
expanded by the orange part
can explain the FNAL+BNL and BNL results  at $2\sigma$ level, respectively. 
The black region is excluded by Xenon-1T $90\%$ CL limits. The region to the right 
of blue dash lines spoils stability of the electroweak vacuum. The grey region gives
an over-abundance for dark matter. The region between the 
ATLAS upper and lower lines is excluded by 13 TeV LHC search 
of slepton pair production at $95\%$ CL.
}
\label{mssm}
\end{figure}

In the low energy effective minimal SUSY model(MSSM) or called phenomenological MSSM(pMSSM).
the masses of bino ($M_1$), winos ($M_2$), higgsinos ($\mu$) and smuons/sneutrino 
($M_{\tilde\ell}$) are all independent parameters. As shown in Fig.~\ref{mssm} \cite{gm2-gutsusy-01}, 
considering other constraints including the dark matter relic density, the dark matter direct detection, 
the vacuum stability and the LHC search for sleptons, the survived MSSM parameter space 
can allow for the explanation of the muon $g-2$ at $2\sigma$ level~\cite{gm2-gutsusy-01,gm2-mssm-01,gm2-mssm-02,gm2-mssm-03,gm2-mssm-04,gm2-mssm-05,gm2-mssm-06,Iwamoto:2021aaf,Gu:2021mjd,Cox:2021gqq}.

\begin{figure}[htb]
\begin{center}
\includegraphics[width=11cm]{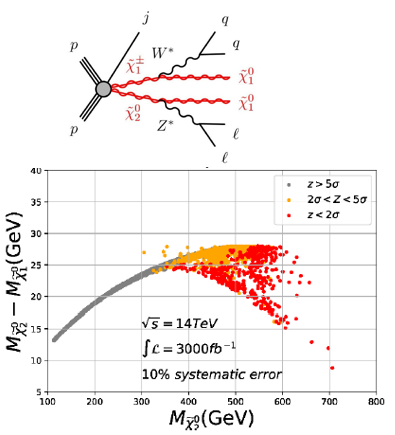}
\end{center}
\vspace{-.5cm}
\caption{ The observability of the MSSM light bino-wino-slepton scenario 
for the explanation of the muon $g-2$ at $2\sigma$ level \cite{gm2-mssm-01}.  
}
\label{bwl-lhc}
\end{figure}

The SUSY explanation of 
 the muon $g-2$ at $2\sigma$ level requires light electroweakinos and sleptons, which can
be most covered at the HL-LHC \cite{gm2-mssm-01,Aboubrahim:2021ily}. For example, in the scenario where the dark matter is 
bino-like with bino-wino coannihilation to achieve the correct dark matter relic density,
  the muon $g-2$ at $2\sigma$ level requires light bino and winos, whose pair production 
at the LHC can be well probed, as shown in Fig.\ref{bwl-lhc}.

\begin{figure}[htb]
\begin{center}
\includegraphics[width=11cm]{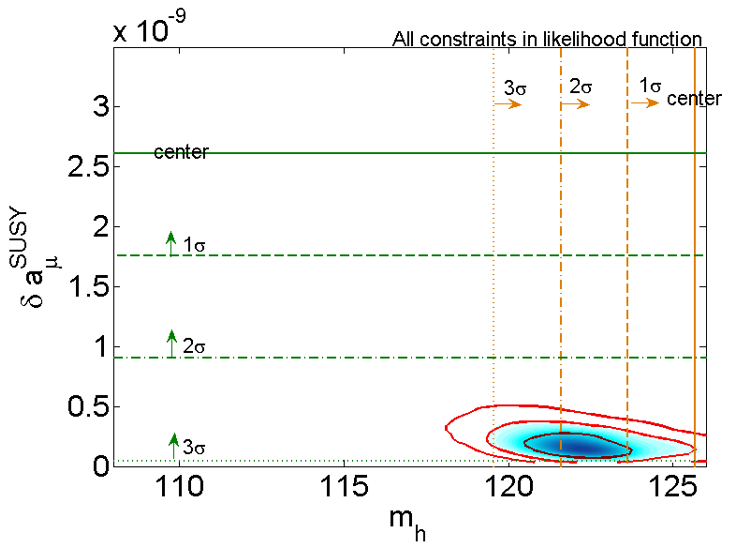}
\end{center}
\vspace{-.5cm}
\caption{ The result of a global fit of the mSUGRA/CMSSM, showing  the tension 
between the 125 GeV SM-like Higgs boson mass and the muon $g-2$ \cite{cmssm-fit}. 
}
\label{tension}
\end{figure}

While the muon $g-2$ anomaly can be readily explained in the low energy effective MSSM due to its large number of free soft SUSY-breaking parameters, the explanation is rather challenging in the constrained models with given SUSY breaking mediation mechanisms ~\cite{gm2-gutsusy-01,gm2-gutsusy-02,gm2-gutsusy-03,Han:2020exx,Lamborn:2021snt,Yin:2021mls}, such as mSUGRA/CMSSM, GMSB and AMSB.   
In these fancy models we have boundary conditions for the soft SUSY-breaking terms at some high 
energy scale, e.g., the GUT scale in mSUGRA/CMSSM. Due to the  boundary conditions, the soft
masses at the weak scale are correlated. To give a 125 GeV SM-like Higgs boson mass, the 
stop masses must be above TeV scale and the correlated slepton masses cannot be as light
as required by the explanation of the muon $g-2$ anomaly. For the mSUGRA/CMSSM, the tension 
between the 125 GeV SM-like Higgs boson mass and the muon $g-2$ is shown in Fig.\ref{tension}.
 In order to accomodate a 125 GeV 
SM-like Higgs boson mass and the muon $g-2$ at $2\sigma$ level, these models need to improved 
or extended. For example, one can make colored sparticles much heavier than uncolored sparticles 
\cite{gm2-SUGRA-ext-01,gm2-SUGRA-ext-02,gm2-SUGRA-ext-03,gm2-SUGRA-ext-04,gm2-GMSB-AMSB-ext-01,gm2-GMSB-AMSB-ext-02} 
(so stops can be much heavier than sleptons and gluino can be much heavier than electroweakinos at
weak scale) or couple the messengers with the Higgs doublets (so the 125 GeV SM-like Higgs boson 
mass can be achieved without too heavy stops) \cite{GMSB-yukawa-higgs}. 

If we go beyond the minimal framework of SUSY, the accomodation of the 125 GeV 
SM-like Higgs boson mass and the muon $g-2$ at $2\sigma$ level may becomes easy.  
For example,  in the next-to-minimal SUSY model (NMSSM) a singlet Higgs field is introduced 
and hence the 125 GeV SM-like Higgs boson mass can be obtained at tree level without the 
large effects of heavy stops (the relatively light stops make this model more natural than
the MSSM \cite{nmssm-mssm}). The explanation of the muon $g-2$ can be achieved in the NMSSM \cite{gm2-nmssm-01,gm2-nmssm-02,gm2-nmssm-03}.     
 
Noe that there have been some attempts to explain the muon $g-2$ 
together with other physical problems (e.g. the proton radius puzzle \cite{Zhu:2021vlz}). Especially, if the value of the fine structure constant measured by the Berkeley experiment~\cite{berkeley} is used, the electron $g-2$ predicted by the SM will also deviate from the experimental value~\cite{Aoyama:2019ryr,Hanneke:2008tm}. A joint explanation of
such an electron $g-2$ anomaly and the muon $g-2$ anomaly is shown to be feasible in SUSY \cite{joint-01,joint-02,joint-03,joint-04,joint-05,Yang:2020bmh}. 
If we further consider the correlation between muon $g-2$ and electron EDM, the CP-phases in SUSY can be stringently constrained \cite{cp-phase-01,cp-phase-02}.  
In this review we do not cover various non-SUSY explanations 
of the $g-2$ or EDM in miscellaneous extensions of the SM, say the 2HDM (see, e.g.,  \cite{Keus:2017ioh,Wang:2016ggf,Wang:2018hnw}), the 3-3-1 model (see, e.g., \cite{Hue:2021zyw,Li:2021poy}) and flavor models (see, e.g., \cite{Calibbi:2020emz,Han:2020dwo,Yin:2021yqy}).        

\section{Outlook} 
\begin{figure}[htb]
\begin{center}
\includegraphics[width=11cm]{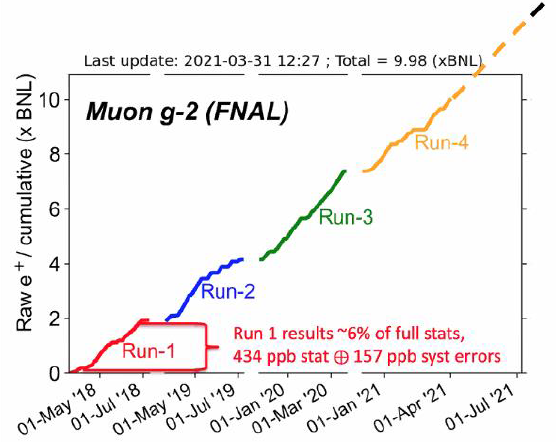}
\end{center}
\vspace{-.5cm}
\caption{ The Fermilab experiment plan on the muon $g-2$ \cite{fnal-data-plan}. 
}
\label{fnal-plan}
\end{figure}

Since the muon $g-2$ may be sensitive to new physics beyond the SM, it will continue playing an
important role in the probe of new physics. So the Fermilab experiment on the measurement of the 
muon $g-2$ has been and will continue to be a focus point in particle physics. So far the result
reported by the Fermilab is rather preliminary and only used $6\%$ data of its planned ultimate 
yield, as shown in Fig.\ref{fnal-plan}. 
In the coming years, Fermilab will continue updating its result amd finally achieve a 
measuement 4 times as precise as the BNL. If the central measured value and the SM prediction are 
not changing much, the deviation shown from the final Fermilab result would reach the $5\sigma$ level.
The forthcoming updated results on both the experimental value and the SM calculation improvement 
(especially from the lattice groups) will bring a huge impact on new physics like the low energy 
supersymmetry.  We just continue to monitor the developments on this front.    

\addcontentsline{toc}{section}{Acknowledgments}
\acknowledgments
This work was supported by the National Natural Science Foundation of China 
(NNSFC) under grant Nos.11821505 and 12075300,  
by Peng-Huan-Wu Theoretical Physics Innovation Center (12047503),
by the CAS Center for Excellence in Particle Physics (CCEPP), 
by the CAS Key Research Program of Frontier Sciences, 
and by a Key R\&D Program of Ministry of Science and Technology of China
under number 2017YFA0402204.

\addcontentsline{toc}{section}{References}
\bibliographystyle{JHEP}
\bibliography{bibliography}

\end{document}